\documentclass[aps,prl,twocolumn,showpacs]{revtex4}

\usepackage{dcolumn}                    
\usepackage{bm}                         
\usepackage{graphicx}

\newcommand {\ba} {\begin{eqnarray}}
\newcommand {\ea} {\end{eqnarray}}

\usepackage{times}

\begin{document}

\title{Shadow gap in the over-doped (Ba$_{1-x}$K$_x$)Fe$_2$As$_2$ compound}

\author{Yunkyu Bang}
\email[]{ykbang@chonnam.ac.kr} \affiliation{Department of Physics,
Chonnam National University, Kwangju 500-757, Republic of Korea}

\begin{abstract}
The electron band around $M$ point in
(Ba$_{1-x}$K$_x$)Fe$_2$As$_2$ compound -- completely lifted above
the Fermi level for $x > 0.7$ and hence has no Fermi Surface (FS)
-- can still form an isotropic s-wave gap ($\Delta_e$) and it is
the main pairing resource generating an s-wave gap ($\Delta_h$)
with an opposite sign on the hole pocket around $\Gamma$ point.
The electron band developing the SC order parameter $\Delta_e$ but
having no FS displays a {\it shadow gap} feature which will be
easily detected by various experimental probes such as
angle-resolved photoemission spectroscopy (ARPES) and scanning
tunneling microscope (STM). Finally, the formation of the nodal
gap $\Delta_{nodal}$ with $A_{1g}$ symmetry on the other hole
pocket with a larger FS is stabilized due to the balance of the
interband pairing interactions from the main hole band gap
$\Delta_h=+\Delta$ and the hidden electron band gap $\Delta_e =
-\Delta$.

\end{abstract}

\pacs{74.20,74.20-z,74.50}

\date{\today}
\maketitle

{\it Introduction:} The superconducting (SC) transition is the
most well known example of the Fermi surface (FS) instability
along with other density wave instabilities such as spin density
wave (SDW), charge density wave (CDW), etc. Mathematically, it is
summarized by a pairing susceptibility $\chi(T) = \lambda
\ln{[\Lambda_{hi}/T]}$ of a conduction band of the Bloch states
\cite{BCS}, where $\lambda$ is a dimensionless coupling constant
and $\Lambda_{hi}$ is the high energy cut-off of the pairing
interaction (for example, $\Lambda_{hi}=\omega_D$, Debye
frequency, for phonon interaction). For the conduction band with a
Fermi surface (FS), the low energy cut-off is in fact zero because
the presence of FS allows the zero energy excitations, which is
now replaced by $T$ at finite temperature in the above formula.
This susceptibility displays the logarithmic divergence with
lowering temperature, hence no matter how weak the pairing
interaction $\lambda$ is, the instability condition, $\chi(T)
\rightarrow 1$, is achieved by decreasing temperature $T
\rightarrow T_c = \Lambda_{hi} \exp{[-1/\lambda]}$. This is called
the FS instability.
However, if there exists a finite low energy cut-off
$\Lambda_{low}$, for example, because there is no FS, then the
susceptibility becomes $\chi \sim \lambda
\ln{[\Lambda_{hi}/\Lambda_{low}]}$ and the instability condition
$\chi(T) \rightarrow 1$ can only be satisfied when the coupling
strength $\lambda$ becomes sufficiently strong, i.e. $ \lambda >
\lambda_{crit}=\frac{1}{\ln{[\Lambda_{hi}/\Lambda_{low}]}}$. This
hypothetical exercise shows that the instability can still occur
with Bloch states without the FS if the coupling is strong enough.
However, notice that the susceptibility $\chi$ becomes temperature
independent in this case, hence this mechanism cannot derive a
phase transition in real system by decreasing temperature.
Therefore, we confirm a common knowledge: {\it no FS, no phase
transition with Bloch states.}

In this paper, however, we demonstrate that the presence of the
low energy cut-off in the pairing susceptibility does not prohibit
the superconducting (SC) phase transition in the multi-band SC
pairing model mediated by an interband pairing interaction as most
probably realized in the Fe-based superconductor
\cite{Kamihara,Stewart}. In particular, in the case of the hole
over-doped (Ba$_{1-x}$K$_x$)Fe$_2$As$_2$ compound, it is known
that the electron band is completely lifted up above the Fermi
level, hence the FS of the electron pocket disappears, for $x >
0.7$ \cite{H Ding}. In this case, we show that the SC order
parameter (OP) should still be formed in the electron band,
which has no FS, as well as in the hole band, maintaining the general
structure of the sign-changing s-wave pairing state mediated by
the antiferromagnetic (AFM) spin fluctuations.

The formation of a SC OP in the band without FS is an
unprecedented SC state and its identification will be a
smoking-gun evidence proving the pairing mechanism of the
Iron-based superconductors mediated by the interband repulsive
interaction\cite{Mazin,Kuroki,other theory,Bang-model}. This SC
gap state without FS will display a {\it shadow gap} feature in
various physical properties and this {\it shadow gap} feature can
be easily detected by ARPES, STM, etc. Finally, the formation of
the SC pairing condensate in the electron band, although it is not
visible at the Fermi level, is the main deriving force to
determine the SC transition temperature $T_c$ and also plays an
important role to stabilize a nodal SC gap in the second and/or
third hole pocket with a larger FS area. Our scenario naturally
explains the $T_c$ variation and the evolution of a nodal gap in
(Ba$_{1-x}$K$_x$)Fe$_2$As$_2$ compound with K doping\cite{H Ding
2,Avci,S Shin, Budko}.

{\it $T_c$ with the electron band lifted above Fermi level:} For
the purpose of demonstration, we start with a minimal two band
model\cite{Bang-model}: one hole band around $\Gamma$ point and
one electron band around $M$ point in the folded Brillouin Zone
(BZ). The pairing interaction is also assumed as a simple
phenomenological form induced by the AFM spin fluctuations defined
as
\begin{equation}
V_{AFM}(k,k^{'}) = V_M \frac{\kappa^2}{|(\vec{k}-\vec{k^{'}})-\vec{Q}|^2
+\kappa^2}.
\end{equation}

\noindent where the AFM correlation wave vector $\vec{Q}$ is
assumed to be $\vec{Q} = (\pi \pm \delta, \pi \pm \delta)$ to
incorporate an icommensurability\cite{Yamada}. The coupled gap
equations are written as

\begin{eqnarray}
\Delta_h (k)  &=&  -  \sum_{k^{'} }  [V_{hh}
(k,k^{'}) \chi_h (k^{'}) \Delta_h (k^{'})
\nonumber \\ & & + ~~V_{he}
(k,k^{'}) \chi_e (k^{'}) \Delta_e (k^{'})], \\
\nonumber \\
\Delta_e (k)  &=&   -  \sum_{k^{'} } [V_{eh}
(k,k^{'})   \chi_h (k^{'})  \Delta_h (k^{'})\nonumber  \\ & & + ~~V_{ee}
(k,k^{'})  \chi_e (k^{'}) \Delta_e (k^{'})]. \\ \nonumber
\end{eqnarray}
\noindent where $V_{hh} (k,k^{'})$, $V_{he} (k,k^{'})$, etc are
the interaction defined in Eq.(1) and the subscripts are written
to clarify the meaning of $V_{hh} (k,k^{'})$ =$V (k_h,k^{'} _h)$,
$V_{he} (k,k^{'})$ =$V (k_h ,k^{'}_e)$, etc., and $k_h$  and $k_e$
specify the momentum $k$  located on the hole and electron bands,
respectively. For the convenience of the analysis of $T_c$, we
introduce the FS averaged pairing potential $<V_{he}
(k,k^{'})>_{FS}=V_{he}=V_{eh}$, then the coupled
$T_c$-equations are written as
\begin{eqnarray}
\Delta_h  &=&   -  \bigl[ V_{hh}N_h \chi_h  \bigr] \Delta_h
 -   \bigl[ V_{he} N_e \chi_e  \bigr]
\Delta_e ,
\\ \nonumber
\Delta_e  &=&   -  \bigl[ V_{ee} N_e \chi_e   \bigr] \Delta_e  -
\bigl[ V_{eh} N_h \chi_h  \bigr] \Delta_h .
\end{eqnarray}
where the pair susceptibility is defined as
\begin{eqnarray}
\chi_{h,e}(T) &=&  \int _0 ^{\Lambda_{hi}} \frac{d \xi}{\xi}
\tanh (\frac{\xi}{2 T}) \approx \ln \Big[\frac{1.14 \Lambda_{hi}}{T}\Big],
\end{eqnarray}
and $N_{h,e}$ are the density of states (DOS) of the hole band and
electron band, respectively.
For simplicity of demonstrating the mechanism, we temporarily drop
the intraband interaction $V_{hh}$ and $V_{ee}$, which are always
much weaker than $V_{he}=V_{he}$. Then the gap equations can be combined
to be
\begin{eqnarray}
\Delta_h  &=&   \bigl[ V_{he} N_e \chi_e  \bigr] \cdot \bigl[ V_{eh} N_h \chi_h  \bigr]
\Delta_h, \\
& \approx & \bigl[ V_{he} N_e V_{eh} N_h \bigr] \Big[ \ln \big(\frac{1.14 \Lambda_{hi}}{T}\big) \Big]^2 \Delta_h
,
\end{eqnarray}

\noindent hence we can read off the critical temperature
\begin{equation}
T_{c0} \approx 1.14 \Lambda_{hi} \exp{\big[-\frac{1}{\lambda_{eff}}\big]}
\end{equation}
\noindent with $\lambda_{eff}=\sqrt{V_{he} N_e V_{eh} N_h}$. Now
if the electron band does not cross the Fermi level and the bottom
of the band is above the Fermi level by $\epsilon_b$,  the only
modification \cite{note 2} of the above analysis is to replace the
susceptibility of the electron band as follows
\begin{equation}
\chi_{e} = \int _{\epsilon_b} ^{\Lambda_{hi}} \frac{d \xi}{\xi}
\tanh (\frac{\xi}{2 T}) \approx \ln \Big[\frac{1.14 \Lambda_{hi}}{\epsilon_b}\Big],
\end{equation}

\noindent so that the coupled susceptibility in Eq.(7) changes
from $\Big[ \ln \big(\frac{1.14 \Lambda_{hi}}{T}\big) \Big]^2$  to
$\Big[ \ln \big(\frac{1.14 \Lambda_{hi}}{T}\big) \Big] \Big[ \ln
\big(\frac{1.14 \Lambda_{hi}}{\epsilon_b}\big) \Big]$ and then,

\begin{equation}
T_c (\epsilon_b) \approx 1.14 \Lambda_{hi} \exp{(-1/\tilde{\lambda}_{eff})}
\end{equation}
\noindent with $\tilde{\lambda}_{eff}=[V_{he} N_e V_{eh} N_h]
\cdot \ln \Big[\frac{1.14 \Lambda_{hi}}{\epsilon_b} \Big]$. Notice
that this analysis is accurate only when $\epsilon_b
> T_{c0}$, that is the same condition as $\ln \Big[\frac{1.14 \Lambda_{hi}}{\epsilon_b}
\Big] < 1$. In the other limit the susceptibility of the
electron band $\chi_e$ of Eq.(9) should be numerically calculated.
Nevertheless the above analysis and Eq.(10) clearly demonstrate the fact that in the multiband
pairing model mediated by the interband interaction the FS
instability still operates even if the FS of the electron band
disappears  and the $T_c (\epsilon_b)$ will only continuously decrease
as the bottom of the electron band $\epsilon_b$ is lifted up above
the Fermi level.

\begin{figure}
\noindent
\includegraphics[width=90mm]{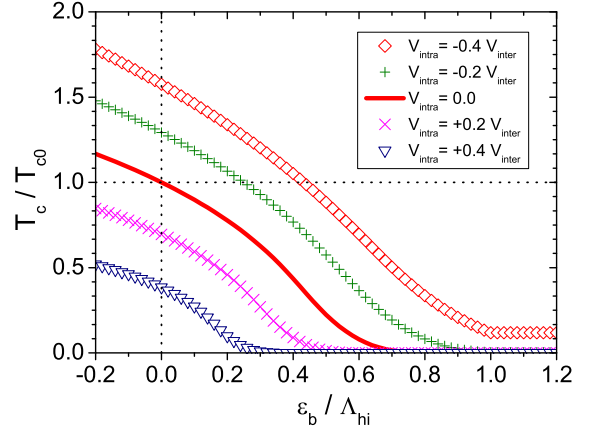}
\caption{(Color online) Numerically calculated $T_c$ of the two band model vs the bottom of the
electron band $\epsilon_b$ for different values of the intra-band interaction
$V_{intra} / V_{inter} = -0.4, -0.2, 0.0, +0.2,$ and $+0.4$. $T_c$ values are normalized by
$T_{c0} (V_{intra}=0.0, \epsilon_b =0.0)$.
We assumed a repulsive interaction for $V_{inter} =V_{he} = V_{eh} >0$
but considered both repulsive and attractive interactions for $V_{intra}$.
For simplicity but without loss of generality, we also chose $N_h = N_e$ and $V_{intra} =V_{ee} = V_{hh}$.
\label{fig1}}
\end{figure}

In Fig.1, we show the numerical results of the exact $T_c (\epsilon_b)$
calculated with Eq.(2) and Eq.(3) including both interband
interaction ($V_{eh}$) and intraband interactions ($V_{ee}$ and
$V_{eh}$). The positive $\epsilon_b > 0$ value is the distance of
the bottom of the electron band above the Fermi level; therefore no
FS exists for the electron band. The negative $\epsilon_b < 0$
value means that the electron band slightly sinks below the Fermi
level, hence has a small FS.
We assumed the repulsive inter-band interaction
$V_{inter}=V_{eh}=V_{he} > 0$ to induce the $\Delta_{\pm}$ gap
solution\cite{Bang-model}. However, for the intra-band
interaction, we considered both repulsive and attractive interaction for generality. 
The attractive intra-band interaction can be possibly
caused by phonons \cite{Bang-phonon} or by the orbital
fluctuations\cite{Kontani}.
The overall behavior of $T_c (\epsilon_b)$ as a function of
$\epsilon_b$ is similar for all cases; linear decrease for small
$\epsilon_b$ value and then exponential decrease for large
$\epsilon_b$ value in accord with Eq.(10). This behavior is
qualitatively in agreement with the $T_c$ variation with K doping
of (Ba$_{1-x}$K$_x$)Fe$_2$As$_2$ compound\cite{H Ding 2,Avci,S Shin,
Budko} and we can understand that the main reason of the
decrease of $T_c$ in (Ba$_{1-x}$K$_x$)Fe$_2$As$_2$ compound is the
lifting of the bottom of the electron band $\epsilon_b$ with K
doping. Furthermore it shows that even when $\epsilon_b$ is lifted
up above the Fermi level and hence the FS of the electron band
completely disappears, the pairing mechanism mediated by the repulsive
interband scattering $V_{inter}$ between the hole and
electron bands continues to operate.
When the intra-band interaction $V_{intra}$ is sufficiently
attractive (the case $V_{intra} = - 0.4 V_{inter}$ in Fig.1), the
$T_c$ finally converges to the limit where the only the hole band
forms a SC transition with the attractive interaction.

{\it Shadow Gap of the electron band:} Here we solved the coupled
gap equations Eq.(2) and Eq.(3) with the realistic tight binding
bands \cite{Bang-model} and the fully momentum dependent
phenomenological pairing interaction of Eq.(1). Although it is not
crucial for the results of this paper in the following, we also
employed the incommensurability ($\delta=0.32 \pi$) of the spin
fluctuations which is recently measured by the neutron experiment
in KFe$_2$As$_2$ by Yamada and coworkers \cite{Yamada}. We used
two model bands: $\epsilon_{h} (k)=t_1 ^h (\cos k_x +\cos k_y) +
t_2 ^h \cos k_x \cos k_y + \epsilon_h$ and $\epsilon_{e} (k)=t_1
^e (\cos k_x +\cos k_y) + t_2 ^e \cos \frac{k_x}{2} \cos
\frac{k_y}{2} + \epsilon_e$ with the band parameters as
(0.30,0.24,-0.6) for hole band and (1.14,0.74,$\epsilon_e$) for
electron band with the notation ($t_1, t_2, \epsilon$). The
electron bands located in the $M$ points ($(\pi, \pi)$ in the
folded BZ) is artificially lifted by shifting the parameter
$\epsilon_e$. For example, we need $\epsilon_e=2.28$ to set
$\epsilon_b =0.0$ which is the case when the bottom of the
electron band just touches the Fermi level. With these parameters,
the DOS of hole band near FS $N_h (0)$ and the DOS of the electron
band near the bottom $N_e (0)$ are $\sim 0.45 /eV$ and $\sim 0.19
/eV$, respectively. Other parameters such as the coupling strength and the pairing cut-off energy $\omega_{AFM}$, etc need not be accurate since all results are renormalized by the gap values.
With the above parameters, the ratio of gap sizes is $|\Delta_{e}|
/ |\Delta_{h}|\sim 3.4$.

\begin{figure}
\noindent
\includegraphics[width=90mm]{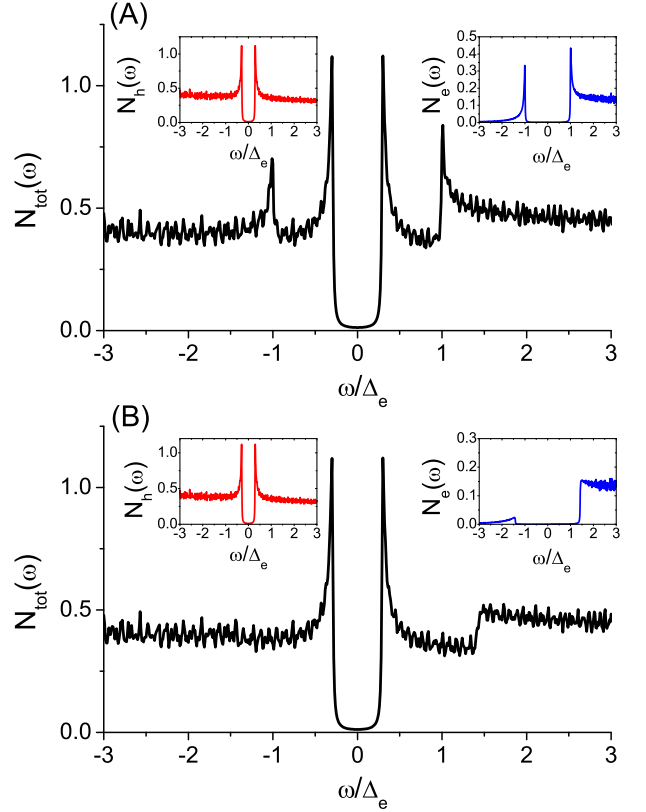}
\caption{(Color online) Calculated density of states, $N_h (\omega)$, $N_e
(\omega)$, and the total $N_{total} (\omega)$ of the two band model in the SC state. (A) is the
case of $\epsilon_b =0.0$ and (B) is the
case of $\epsilon_b =|\Delta_e|$.  \label{fig2}}
\end{figure}

In Fig.2, we show the calculated DOSs, $N_h (\omega)$, $N_e
(\omega)$, and the total $N_{total} (\omega)$. Figure 2.(A) is the
case when $\epsilon_b =0.0$. The electron band exists only above
the Fermi level. Nevertheless, in the SC state, the Bogoliubov
quasiparticles are formed above and below the Fermi level, hence
the DOS $N_e (\omega)$ is created both for $\omega > 0$ and for
$\omega <0$. However, the shape of $N_e (\omega)$ is very
asymmetric for above and below $\omega = 0$ as seen in the right
inset of Fig.2(A). It should be contrasted with $N_h (\omega)$ in
the left inset which is symmetric as a typical SC DOS. The total
DOS $N_{total} (\omega)$ displays this clear signature of the
asymmetric DOS due to the {\it shadow gap} formed in the electron band above the Fermi level.
Fig.2(B) is the case $\epsilon_b =|\Delta_e|$. In this case the
gap size in $N_e (\omega)$ becomes $\sqrt{\epsilon_b ^2 + \Delta_e
^2}$ and the shapes of $N_e (\omega)$ and $N_{tot} (\omega)$
become even more asymmetric than the $\epsilon_b =0$ case. This
predicted asymmetric DOS should be easily detected by the STM
measurement.

\begin{figure}
\noindent
\includegraphics[width=90mm]{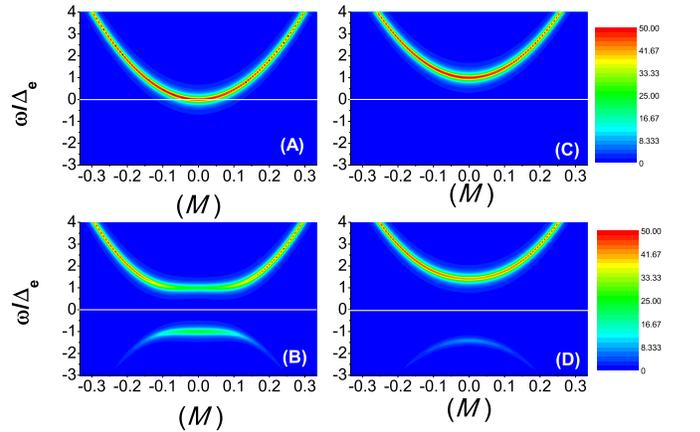}
\caption{(Color online) The quasiparticle spectral density of the electron band
$\epsilon_{e}(k)$ near the $M$ point. (A) and (B) are the normal state and the SC state
of the case $\epsilon_b =0.0$, respectively. (C) and (D) are the
corresponding results of the case $\epsilon_b = |\Delta_e|$. \label{fig3}}
\end{figure}

In Fig.3, we showed the one particle spectral density of the
electron band near the $M$ point. This is calculated by
$\frac{1}{\pi}Im G_e(\omega,k) = \frac{1}{\pi}Im
\frac{\omega+\epsilon_{e}(k)}{\omega^2 -[\epsilon_{e}^2(k) +
\Delta_e ^2]}$. These results are another manifestation of the
shadow gap feature of the electron band which does not have the
FS. Fig.3(A) and (B) are the normal state and the SC state of the
case $\epsilon_b =0.0$, respectively, and Fig.3(C) and (D) are the
corresponding results of the case $\epsilon_b =|\Delta_e|$. In the
normal state, the quasiparticles do not appear below the Fermi
level simply because the band does not exist there. However, when
temperature decreases below $T_c$, the quasiparticle spectral
density appears below the Fermi level. This dramatic effect should
be easy to be detected by the ARPES measurement and in fact it
seems already detected in other Iron-based superconducting
compound FeTe$_{0.6}$Se$_{0.4}$ by Shin and coworkers
\cite{BCS-BEC} although the interpretation of the authors
\cite{BCS-BEC} is somewhat different than ours.

\begin{figure}
\noindent
\includegraphics[width=90mm]{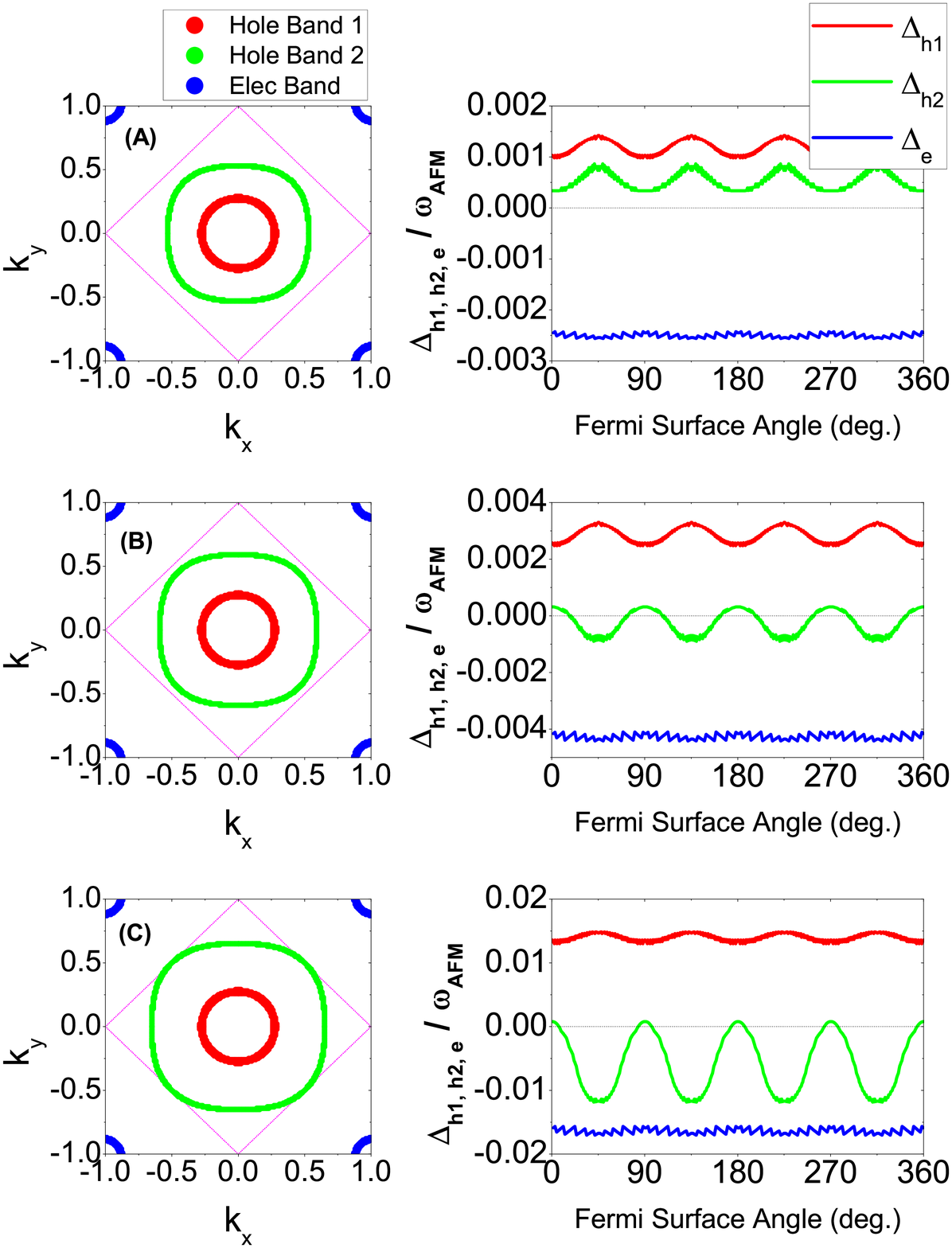}
\caption{(Color online) (Left panel) The FS evolution of three band model; only the $h2$-band (green line)
varies and the $e$-band FS is only for showing and in real calculations it has zero size with $\epsilon_b =0$.
(Right panel) The corresponding gap solutions $\Delta_{h1}$ (red line), $\Delta_{h2}$ (green line),
$\Delta_{e}$ (blue line), respectively. \label{fig4}}
\end{figure}

{\it Evolution of nodal gap in (Ba$_{1-x}$K$_x$)Fe$_2$As$_2$}: The
end member of (Ba$_{1-x}$K$_x$)Fe$_2$As$_2$ compound, KFe$_2$As$_2$, has
been considered as the most strong candidate for a nodal gap
superconductor among the Iron-based superconductors
\cite{K122-nodal}. And the optimal doped
Ba$_{0.6}$K$_{0.4}$Fe$_2$As$_2$ is well confirmed to
have the isotropic full s-wave gaps \cite{Stewart,optimal}.
Therefore the evolution from a full gap to a nodal gap in
(Ba$_{1-x}$K$_x$)Fe$_2$As$_2$ compound has been a keen interest in
the past years \cite{H Ding 2,Avci,S Shin, Budko}.

In this section, in order to study the gap evolution in
(Ba$_{1-x}$K$_x$)Fe$_2$As$_2$, we introduce the minimal three band
model. In particular, we focus on the relation between the FS size
and the anisotropic gap or nodal gap evolution. We add one more
hole band $\epsilon_{h2} (k)=t_1 ^{h2} (\cos k_x +\cos k_y) + t_2
^{h2} \cos k_x \cos k_y + t_3 ^{h3} (\cos 2k_x + \cos 2k_y) +
\epsilon_{h2}$ to the previous studied two band model, so that we
have two hole bands $h1$ and $h2$ and one electron band $e$. The
second hole band $h2$ is tuned to have a larger FS than the one of
$h1$-band; we used parameters $(0.7, -0.1, 0.3,
\epsilon_{h2})$ and varied $\epsilon_{h2}$ to change the FS size.
For systematic study of the FS evolution, we fix the sizes of the FS of $h1$-band and
$e$-band. In the case of $e$-band, in fact, we chose to have
$\epsilon_b = 0$, i.e. the FS size of $e$-band is zero. The spin
fluctuation interaction $V_{AFM}({\bf q})$ given by Eq.(1) is also
fixed with $\kappa = 0.3 \pi$ and $\delta=0.32 \pi$.

In Fig.4, we showed the gap solutions for the several different
size of the $h2$-hole pocket. In the left panel, the FSs of three
bands are drawn for different values of $\epsilon_{h2}$ in the
folded BZ. As said above, $h1$-band and $e$-band are fixed while
only the FS size of $h2$-band varies. Also the $e$-band pocket is
drawn only for showing but in real calculations its size is zero
because we chose $\epsilon_b = 0$.
As expected, when the $h2$-hole pocket is close to the
$h1$-hole pocket as in Fig.4(A), the gap solution is basically a
$\pm$s-wave state: the hole bands have all
$+\Delta$ and electron band has all $-\Delta$ despite some degree of 
anisotropy. The degree of anisotropy is determined by the
sharpness of the pairing interaction $V_{AFM}({\bf q})$ in
momentum space, which is determined by $\kappa \sim \xi^{-1}$, and
the FS sizes. The reason why the average sign of $h2$-band gap is
"$+$" is because the distance between $h2$-band pocket and $e$-band pocket in
the BZ is closer to $(\pi \pm\delta, \pi \pm\delta)$ than what the
distance between $h2$-band pocket and $h1$-band pocket is to $(\pi \pm\delta,
\pi \pm\delta)$.

With increasing the $h2$-pocket, the $h2$-band gap $\Delta_{h2}$
obtains the "$+$" section and "$-$" section, hence becomes a nodal
gap with $A_{1g}$ symmetry. 
In our simple toy model with a simple phenomenological interaction Eq.(1) 
between all bands -- both intra and inter -- and without any
orbital degrees of freedom, this dramatic gap evolution only with
a small increase of the FS size, fixing all other parameters, is
demonstrating that the subtle balance of the repulsive and
attractive interactions between bands is the crucial mechanism to
induce and stabilize the nodal gap solution. In the case of
Fig.4(B), both $+\Delta$ gap of the $h1$-band and $-\Delta$ gap of
the $e$-band exert a similar strength of the repulsive and attractive interactions, respectively, to the $h2$-band from the same $V_{AFM}({\bf q})$. Therefore the $h2$-band
should maximize the condensation energy gain by properly distributing $+$ OP and
$-$ OP, hence developing accidental nodes but keeping $A_{1g}$
symmetry because of the crystal symmetry. Further increasing the
$h2$-pocket size in Fig.4(C), the average repulsive interaction from the $h1$-band is larger than the average attractive interaction from the $e$-band, hence $\Delta_{h2}$ develops more negative lobes;
here the average interactions $<V_{AFM}({\bf
q})>_{h1,h2}$ and $<V_{AFM}({\bf q})>_{e,h2}$ depend on the weighting of DOSs $N_{h1}$, $N_{h2}$ and
$N_{e}$ and the average ${\bf q}$ value in comparison to $(\pi \pm\delta, \pi \pm\delta)$.

Assuming that the $h2$-pocket size increases in (Ba$_{1-x}$K$_x$)Fe$_2$As$_2$ 
with the K doping, the overall gap anisotropy of the $h2$-band and the systematic
development of the nodal gap structure shown in Fig.4 is surprisingly consistent
with the recent ARPES observation by Shin and
coworkers\cite{Shin-2}. Our $h2$-band should be compared to the
outer hole band and our $h1$-band represents the inner and middle
bands in Ref.\cite{Shin-2}. Of course, there is a discrepancy, that
is, the overall gap size(s) increases with the $h2$-pocket size in
our calculations which is clearly opposite to the experimental
observation. However, in our calculations in Fig.4, we fixed the
$\epsilon_b$ and $V_{AFM}({\bf q})$ both of which should change
with K doping in (Ba$_{1-x}$K$_x$)Fe$_2$As$_2$ toward the
direction reducing $T_c$ and the gap sizes, so that this discrepancy can naturally
be cured with a more realistic model.

{\it Conclusion}: We showed that the absence of the FS does not
ruin the FS instability of the SC transition in the multi-band
pairing model mediated by the interband scattering. We
demonstrated that the $T_c$ evolution with the bottom of the
electron band $\epsilon_b$ can consistently explain the
experimental $T_c$ evolution of (Ba$_{1-x}$K$_x$)Fe$_2$As$_2$
compound. As a smoking-gun evidence of this hidden band pairing proposed in this paper, we
predicted the {\it shadow gap} features both in ARPES and STM
measurements. Finally, we demonstrated that the hidden electron
band should continue to play an crucial role for the pairing
mechanism as well as the nodal gap development in
(Ba$_{1-x}$K$_x$)Fe$_2$As$_2$ compound.

{\it Acknowledgement -- }This work was supported by Grants No.
NRF-2011-0017079 funded by the National Research Foundation of
Korea.


\begin{references}
\bibitem{note 2}
It is a good approximation that the DOS $N_e$ near the bottom of
the electron band is a constant because the band has a quasi two
dimensional parabolic dispersion.
\end{references}
\end{document}